# Predictive validities: figures of merit or veils of deception?

*Peter H. Schonemann*
*Purdue University, West Lafayette, USA*
*pschonemann@gmail.com*

*Moritz Heene*
*Ludwig Maximilian University, Munich, Germany*
*heene@psy.lmu.de*

Abstract

The ETS has recently released new estimates of validities of the GRE for predicting cumulative graduate GPA. They average in the middle thirties - twice as high as those previously reported by a number of independent investigators.

It is shown in the first part of this paper that this unexpected finding can be traced to a flawed methodology that tends to inflate multiple correlation estimates, especially those of populations values near zero.

Secondly, the issue of upward corrections of validity estimates for restriction of range is taken up. It is shown that they depend on assumptions that are rarely met by the data.

Finally, it is argued more generally that conventional test theory, which is couched in terms of correlations and variances, is not only unnecessarily abstract but, more importantly, incomplete, since the practical utility of a test does not only depend on its validity, but also on base-rates and admission quotas. A more direct and conclusive method for gauging the utility of a test involves misclassification rates, and entirely dispenses with questionable assumptions and post-hoc "corrections".

On applying this approach to the GRE, it emerges (1) that the GRE discriminates against ethnic and economic minorities, and (2) that it often produces more erroneous decisions than a purely random admissions policy would.

*1. Introduction*

*The GRE*

The Graduate Record Examination (GRE), originally developed by Ben Wood for the Carnegie Foundation in 1935 (Lemann, 1999, p. 35) underwent numerous revisions over the years. The version treated in the GRE report by Burton and Wang (2005) consists of a General Test and a number of subject (achievement) tests. The focus of Burton and Wang (2005) is on the General Test which resembles the Scholastic Aptitude Test (SAT) and consists of three subtests: (V), Quantitative (Q), and Analytical (A). According to a GRE Guide (1993, p. 9), these scores "are related to success at the graduate level of education".

The paper version of the General Test takes 3 ½ hours to complete. The fee for taking the exam was $135 in 2007.

*The GRE Report*

The GRE Board recently published new validities of the GRE for predicting long-term criteria, in particular the cumulative graduate school GPA (Burton & Wang, 2005).

As in the GRE Report, the term "(predictive) validity" is used here in the standard sense defined by Lord and Novick (1968, p. 277), as the "multiple correlation between a set of predictors and a criterion". By this definition, a test has as many validities as criteria one may wish to predict with it. In particular, one distinguishes between short-range, proximal criteria, such as first year GPA, and long-range, distal criteria, such as cumulative grade point average.

The main results extracted from Tables C1, C2, and C6 of this report are summarized for the convenience of the reader in Table 1 of this paper.

==================
TABLE 1 ABOUT HERE
==================

As shown, the study encompassed 5 disciplines: Biology, Chemistry, Education, English, and Psychology. The number of departments in each discipline varied from 2 to 5. The sub-sample sizes (SSS) varied from 10 to 453, with a median around 45. Sixteen out of nineteen 19 (84%) sub-sample sizes were smaller than 100, and 12 (63%) were smaller than 50.

For each department, the uncorrected multiple correlation R and the (restriction of range) corrected value $R^c$ are given for three predictors V (Verbal), Q (Quantitative) , and U (Undergraduate GPA) and their combinations. The criterion of interest is cumulative graduate GPA. The last three columns show the pooled values for U, V+Q and U+V+Q. The raw correlations were not corrected for shrinkage.

*Main Findings of the Report*

The validities for V+Q average .35 *before* correction for restriction of range. This value is almost as high as that for predicting freshman GPA from the SAT, where the criterion is proximal. Moreover, since the applicant pool for the SAT is broader than for the GRE, range restriction should be less severe (Donlon, 1984, p.162f).

*Findings by Previous Independent Investigators*

As Sternberg and Williams (1997) have noted, investigators have repeatedly reported GRE validities in the 20s for *short-term* criteria, such as first year graduate school GPA (Wilson, 1979; Burton & Turner; 1983; Schneider & Briel, 1990), as did, more recently, Morrison and Morrison (1995, p. 311). For various long-term criteria, Sternberg and Williams (1997) found that GRE validities ranged in the teens (Their Table 5, p. 639). Those reported in Horn and Hofer (undated) for graduation were effectively zero (Tables 5, p. 17, and Table 22, p. 82).

*Proximal versus Distal Criteria*

Just as it is harder to predict the weather two weeks ahead than two days ahead, one expects smaller validities for distal than for proximal criteria, such as first year graduate GPA. Humphreys (1968) demonstrated this in a now classic validity study of the ACT, which predicted first semester undergraduate GPA with a validity of .48. As the prediction interval lengthened, the validities steadily declined until they reached .16 for the 8'th semester (p. 376). There is no obvious reason why this effect should not also apply to the quite similar GRE.

*GRE versus GPA*

Equally unexpected is the new finding that V+Q, with only one exception, now outperforms undergraduate GPA, U. In the past, undergraduate GPA usually exceeded V+Q+A by about .05 (e.g., GRE 1992-93 Guide, p. 45).

*Central Hypothesis*

As its title indicates, the new GRE Report is specifically concerned with long-term criteria. It is, therefore, puzzling why most of the *uncorrected* validities for graduate GPA are roughly 15 correlation points higher than those reported by most previous investigators. Only 4 out of 36 (11%) of the values in Table C6 fall into the typical range in earlier studies below .25. Although the sample is small, the small overlap is nevertheless striking.

Thus, the elevated validities in the recent GRE Report not only conflict with previous findings, they also defy common sense, because U tends to be more stable than V+Q+A.

***The central thesis of the present paper is that the elevated GRE validities for long-term criteria can be traced to the so-called "Method of Pooled Department Analysis" (PDA),***

which the authors define on p. 11 of their report:

"Summaries of correlations are averages of the individual department coefficients corrected for multivariate restriction of range and weighted by the number of students in the department."

The uncorrected validities, which are also given, are of primary interest here. Corrections will be taken up in sec. 5.

*Pooling versus Aggregating Sub-sample Data*

In principle, there are two different methods for combining data from sub-samples:

1. Aggregate the *scores* of the smaller samples into a larger sample, or
2. Compute the desired *statistics* (in the present case, correlations) in each of the sub-samples, and then average them to obtain parameter estimates for the whole data set.

Method 1 is the conventional procedure. Method 2 is the Method of Pooled Department Analysis (PDA) used in the new Report. For a numerical illustration, see Table 1B.

*Research Strategy*

Section 3 describes the results of computer simulations confirming the Central Hypothesis that the elevated validities in the GRE Report are due to the PDA type of data pooling, because it increases the upward bias inherent in multiple correlation estimates.

The basic research strategy was to construct a number of population correlation matrices with specified validities, and then to draw samples from which the population parameters are estimated using various pooling methods.

## 2. Technical Aspects of the Simulations

Readers not interested in the technical details of the simulations may move on to the Results section 3.

A computer program was written with IMP (Schonemann, Schuboe & Haagen, 1988) to generate multivariate normal random variables with specified population correlation matrix $\Sigma$, as follows:

For a given number of variables p and number of replications SST, SST$\times p$ uncorrelated random numbers were drawn from a *p*-variate $N(\Phi, I)$ distribution and assembled in an SST$\times p$ score matrix *X*. The given correlation matrix $\Sigma$ of order $p \times p$ was then factored into a gram product, $\Sigma = AA'$. The product $Y = XA'$ is then an SST-fold sample from an $N(0, \Sigma)$ distribution (since $E(Y) = \Phi A' = \Phi$ and $E(Y'Y) = AE(X'X)A' = AA'$, and the y's are linear combinations of the *x*'s).

This total sample of size SST was then partitioned into NSS sub-samples of equal sub-sample size SSS, so that NSS×SSS = SST. Concretely, if the total sample size was SST = 1000, and the sub-sample sizes were SSS = 50, then the program generated NSS = 20 sub-samples of size 50.

For each sub-sample, the p×p correlation matrix R was computed to obtain p multiple correlations using the formula

$R_i = [1 - 1/r^{ii}]^{1/2}$,

where $r^{ii}$ denotes the *i*'th diagonal element of the inverse correlation matrix. This formula follows from the expression for inverses of partitioned matrices (e.g., Searle, 1966, p. 210). Guttman (1953) introduced it into Psychology in his work on image analysis. Besides being computationally efficient, it has the added advantage of producing all p multiple correlations simultaneously, with each variable serving in turn as criterion. The resulting NSS row vectors of the within-sample correlation estimates were then averaged to obtain estimates according to the PDA method used in the Report.

Finally, on setting NSS = 1 and SSS = SST, and applying the same computations to the total sample matrix of order SST×*p*, the program produced multiple correlation estimates obtained with the more conventional pooling method 1. In this case, it also computes the sum of the predictors as an alternative decision variable.

These simulations closely follow the precedent set in the Burton and Wang report which, for the most part, was limited to the same regression within each discipline, to justify averaging of the sample correlations across departments in the same discipline.

*3. Simulation Results*

The main results of the simulations are summarized in Tables 2 to 4. Each Table shows the population matrices that were used, followed by the multiple correlations and bias differences for various combinations of sub-sample sizes with number of sub-samples. These parameters were chosen so that the total pooled sample size SST was close to 1000. For the 13×77 combination, it was 1001.

For the present purposes, the correlations in the third (boldface) column of Tables 2 and 3 are of primary interest, since they represent the case when the third variable serves as criterion. The other two columns provide additional bias information for data with higher validities.

=================
TABLE 2 ABOUT HERE
=================

*General case (Table 2)*

The population correlation matrix shown in Table 2 was meant to represent the typical case, in which the two predictors (variables 1 and 2) correlate more highly with each other than with the criterion (variable 3), and the predictor validities are unequal.

The blocks in the tables are ordered by sub-sample sizes, increasing towards the bottom.

The first row of each block in Table 2, labeled *pop,* shows the population multiple correlations. If variable 3 serves as criterion, it is .301. If variables 2 or 3 serve as criterion, it is .600 or .627, respectively.

The second row in each block, labeled *pda*, gives the correlations obtained on averaging

Predictive validities: figures of merit or veils of deception? 10the NSS within-sub-sample estimates. This is the PDA method of pooling used in the GRE Board Report. The third row, labeled *agr*, gives the correlations that result on aggregating the sub-sample *scores* in the conventional way, resulting in a pooled sample of size SST. The fourth row, labeled *sum,* gives the correlations that result on adding the standardized predictors. Thus, agr is a weighted average, while sum is proportional to the unweighted average in the total sample.

The next three rows give the bias differences, as indicated by the row labels. Bias is defined here as

$b :=$ population parameter – sample estimate.

When this difference is negative, the sample estimate exceeds the population value, which is usually the case.

In Table 2, all bias estimates for variable 3 are negative. As expected, they decrease in magnitude with increasing sub-sample size. The smallest (best) values are tagged with one asterisk, the largest (worst) with two asterisks.

To summarize, inspection of Table 2 shows the following:

1. For variable 3 as the criterion, bias is largest for the PDA method in all three cases;
2. Pooling scores, agr, is uniformly superior to averaging estimates, pda;
3. The magnitude of this superiority effect declines with increasing sub-sample size;
4. For the data in Table 2, bias is smallest for sum;
5. The bias tends to decrease with increasing sub-sample size;
6. The bias tends to increase as the population correlations decrease (rows from left to right).
    This point is important because, in practice, long-term predictive validities tend to be small.

==================
TABLE 3 ABOUT HERE
==================

*Zero validities (Table 3)*

Table 3 is organized in the same manner as Table 2. It serves as a check on the findings reported so far. The population matrix shown in Table 3 differs from that in Table 2 only in that both predictor validities are now set to zero, so that the population multiple correlation is also zero. This permits an assessment of type 1 errors.

All trends reported for Table 2 hold up for Table 3. In particular, the PDA method consistently produced the most biased estimates.

For the smaller sub-sample sizes, 25 and 50, the magnitude of the bias is substantial for pda: -.270 and -.163, respectively. It is only -.040 and -.025 for agr, and -.022 and -.001 for sum. For the largest sub-sample size, 77, the bias of pda is still 5 times larger than the bias of agr.

The results in Table 3 confirm and strengthen those in Table 2. They also graphically illustrate that the bias is substantial when the population validity is actually zero.

Counting asterisks in Tables 2 to 3, one finds that pda placed worst, the simpler sum best, and the traditional pooling method intermediate in all cases. Quantitatively, agr is much closer to sum than to pda. Thus, the PDA method tends to overestimate population validities for small sub-samples (< 79) more severely than the conventional method of simply pooling scores.

==================
TABLE 4 ABOUT HERE
==================

*Bias as a function of validity (Table 4)*

As noted, on scanning the rows in Tables 2 and 3, one finds that the bias seems to vary inversely with the test validity. To follow up on this observation more systematically, Table 4 summarizes the results of a bias comparison across validities.

Section 1 exhibits the population correlation matrices. The third variable serves as the criterion in each case. The correlation between the two predictors was fixed at .6. The individual predictor validities were chosen so as to achieve multiple correlations (validities) as close as possible to the successive values .0, .1, .2, .3, and .4. The actually achieved validities are shown below the correlation matrices.

Section 2 of Table 4 displays the bias values for various numbers of sub-sample (NSS) by sub-sample size (SSS) combinations as before. The third row of each validity block shows the bias of sum. The results for sum are less clear-cut than in Tables 2 and 3. Nevertheless, this simple, presumably widely used scoring method seems to hold up remarkably well against pda.

The last row of Table 4 gives the column averages of the *absolute values* of the bias entries, the last column the corresponding row averages.

These (boldfaced) row- and column averages contain the essential information in Table 4. In particular, one finds

1. From the last row labeled c*mns*: The average magnitude of the bias increases as sub-sample size decreases as was already apparent in Tables 2 and 3;
2. From the column labeled *diff*: Across all validities shown, the traditional score pooling method agr produces uniformly smaller bias than pda;

3. As in Tables 2 and 3, this superiority of agr over pda diminishes with increasing sub-sample size;

4. From the last column labeled *rmns*: Bias increases as population validities decrease.

Observations 1 and 4 are especially relevant to the results in the GRE Report. Observation 1 because 15 out of 19 (79%) of the sub-samples were smaller than 77 (Table C2, p. 56). Observation 4 is important because previous investigators have found that GRE validities for long-term criteria are small. The combination of both factors 1 and 4 in the Report probably rendered the data particularly vulnerable to the increased bias induced by the PDA method of pooling.

To illustrate the bias effect concretely: From Table 4 one learns that when the population validity is exactly zero and the (average) sub-sample size is 25, then the PDA method overestimates it as .27; while agr estimates it as .04. For the larger sub-sample size of 50, pda overestimates the zero correlation as .16, agr as .03.

In the non-null case, if the population validity is .11 and SSS = 25, then pda overestimates it as .17; while agr estimates it correctly as .11. Correlations of .20 are overestimated by pda as .29 when SSS = 25, and as .23 when SSS = 50. The corresponding figures for agr are .18 and .17. Thus, while the agr estimates are not perfect, the overall trend is clear: The PDA method systematically overestimates small validities by a wide margin.

*4. Discussion of the Simulation Results*

The conclusion appears inescapable that the elevated GRE validities presented in the GRE Report are result from the bias-enhancing pooling method PDA.

Intuitively, the reason for the persistent upward bias of the PDA method of pooling is not hard to find and most easily understood on inspecting the regression weights in Table 1C (Table C2 in the Report). They are highly volatile across departments within the same school, bearing little resemblance to each other and, hence, neither to the underlying populations weights. A few examples may suffice to make the point: For the two chemistry departments, they are .213 and -.069 for V, and -.013 and .389 for Q (cf. Table 1C). Thus, in one chemistry department, V is the best sole predictor of cumulative GPA, while in the other, it is Q. Similarly, the weights for the 5 English departments vary between -.001 and 1.330 for V, and between -.318 and .256 for Q.

Within each sub-sample, the PDA method squeezes out the highest possible predictability and summarizes it in a sample multiple correlation. The smaller the sub-sample size, the more the multiple correlations – not just the regression weights - will capitalize on chance. Averaging them across sub-samples will thus yield an exaggerated and misleading estimate of the multiple correlation in the population. The upward bias of the PDA method is most pronounced near the lower bound of the validity range, because multiple correlations cannot be negative by definition. This fact renders the bias distribution J-shaped, bunching up near zero. Unfortunately, as most previous investigators have found, it is precisely this range that is relevant for long-term predictions of the GRE.

As the results for sum in Tables 1 to 3 further show, more reliable estimates could have been obtained for unweighted sums of the predictors than for sums weighted with multiple regression weights. This strategy would have had the added benefit of reflecting what most test users actually do.

*Compatibility with Previous Studies*

The magnitude of the bias agrees with the outcomes of previous GRE validity studies cited earlier. The average bias in the upper left hand corner of Table 3 (for validities below .3 and sub-sample sizes below 51), is .13. If the inflated GRE validity estimates in the GRE Report are discounted by this magnitude, they fall into the interval below .25 covering previously reported validity estimates with a median value in the teens, just as Sternberg and Williams (1997) found. This is relevant because some critics of their results raised concerns that Yale may be atypical of graduate schools in general because it is a highly selective school.

Sternberg and Williams' validities average .22 for the proximal $1^{st}$ year graduate GPA. For long-term criteria such as $2^{nd}$ year GPA, just as in Humphreys' study, the average drops to .03. For the more stable cumulative graduate GPA, Sternberg and Williams found .15 (Table 5, p. 639), a value half as large as those reported in the GRE Report. Thus, the recent GRE Report exaggerates the proportion of explained variance roughly by a factor of 5 (from 2 to 10 percent).

*5. Corrections for Restriction of Range*

Many investigators, the authors of the GRE Report included, point to restriction of range as a reason why long-term validities of the GRE tend to be low: "Restriction in range lowers correlation coefficients, so grade point average will look like a poorer predictor of graduate school outcomes than it really is" (Burton & Wang, p. 10). Some commentators have chastised Sternberg and Williams (1997) for not paying sufficient attention to this problem. Following this logic, the authors of the GRE Report also report their validity coefficients after correcting them upward for restriction of range. In some cases, this transformation raises them dramatically, for example from .24 to .38 or from .25 to .50 (Table C6, p. 60, partially reproduced here in Table 1).

Though widely embraced, this logic can be questioned on several grounds. From a pragmatic point of view, it seems safe to say that, if a test performs poorly in the validation sample, then it will probably also perform poorly in predicting academic success in the subpopulation of applicants to graduate school. The problem, some may argue, is not so much that the uncorrected validities make the test "look like a poorer predictor of graduate school outcomes than it really is", but that the corrected validities make it look better than it really is when it comes to making decisions.

*Systematic Preference for Upward over Downward Corrections*

These concerns deepen once one notices the consistent preference for upward corrections, such as for restriction of range, over downward corrections, such as corrections for shrinkage in ETS reports. Although the authors of the GRE Report briefly touch on the shrinkage problem, they quickly dismiss it again with the disarming argument that "The shrinkage adjustment did not seem compatible with our correction for restriction of range" (p. 11). They found that "The correction for shrinkage reduced the estimated $R^2$ [!] to -.27 (Ed. Dept 1) and -.05 (Eng. Dept 2)" (p. 35, exclamation mark added).

*Necessary Assumptions for Restriction of Range Corrections*

Such absurd negative estimates of squares are a reminder that all such "corrections" depend on assumptions that are rarely tested. If they are violated, then the intended improvements in the estimates are, of course, illusory.

Restriction of range corrections require (1) independence of errors, (2) linear regression, (3) homogeneity of error variances, and (4), for statistical tests, normality. Lord and Novick (1968, p. 233) cite Lord (1960), who tested these assumptions. They concluded:

"1. The distribution of the errors of measurement was not independent of true score.
2. In the high-ability group, error variances decreased on average as true score increased; in the low-ability group, error variances increased on the average as true score increased.
3. Errors of measurements were not normally distributed.
4. In the high ability group, the distribution of the errors of measurement was negatively skewed; in the low-ability group, this distribution was positively skewed.
… The conclusions seem reasonable, or can be plausibly rationalized, as consequences of a floor effect and a ceiling effect … Some similar conclusions were obtained empirically by Mollenkopf (1949)".

Hence, as long as these assumptions remain untested, results of restriction of range corrections remain untrustworthy (Gulliksen, 1950, p. 131f., Lord & Novick, 143f.) and violations of these assumptions can lead to positively biased validity estimates (Brown, Stout, Dalessio & Crosby, 1988).

*Estimating Validities without Need for Corrections*

It deserves mention at this point that, even within the narrow constraints of classical test theory, there is a more direct, less problematic method for estimating the validities of the SAT or GRE for a given applicant pool. All that is required is to give the test to a random sample from the applicant pool and grant all sampled applicants access to higher education regardless of how they scored on the test. After one year, one thus obtains a complete sample for this applicant pool that includes the needed criterion scores to obtain trustworthy validity estimates.

Since the GRE has been in existence for over 70 years it remains a mystery why during all this time no-one seems to have thought of this straightforward method for estimating validities without any need for post-hoc corrections. One might argue that this approach would raise ethical questions about who would be favoured by fortune to be selected. However, such questions are always subject of admission testing. For example, with respect to the Scholastic Aptitude Test (SAT), Crouse and Trusheim (1988) found out: "The SAT therefore acts with respect to admission outcomes much as a zero-validity supplement to high school rank that increases rejection of low-income students" (p. 131). Thus, ethical questions are not restricted to the proposed method but are also an inevitable problem for an admission test *due to* of its possibly low validity.

### 6. Classification Rates as a Superior Method for Appraising the Utility of Tests

Over 50 years ago, Meehl and Rosen (1955) have drawn attention to the fact that the validity of a test alone, no matter how high, never suffices to gauge its effectiveness for making actual decisions, since this depends also on base-rates and admission quotas. This elementary fact is rarely mentioned in ETS reports, including Burton and Wang (2005).

More recently, Schonemann and Thompson (1996) and Schonemann (1997) have expanded on this approach and stressed that it renders the limitations of tests as decision instruments much more transparent than the conventional regression approach that often obscures rather than illuminates the limitations of tests. While non-experts may have difficulty visualizing variances and appraising correlations, they are usually quite at adept at interpreting misclassification rates, which are at the heart of this alternative approach.

A simple and intuitively transparent way for evaluating the merit of a test is to view it as a decision tool. Table 5 defines the basic terms useful for interpreting results obtained with this approach.

=================
TABLE 5 ABOUT HERE
=================

As laid out in Table 5A, one is faced with two given outcomes: The candidate is either qualified or unqualified by some objectively defined criterion (e.g., he/she may either graduate or not graduate). The decision is based on two test outcomes: The candidate either passes or fails the cut-off required for admission. A 2×2 table of joint proportions completely summarizes this scenario. The cell entries in Table 5A have self-explanatory names.

The diagonal cells, tp and tn, represent correct classifications, and the off-diagonal cells, fp and fn, represent misclassifications. The proportion of all candidates passing the test is the (admission-) "quota". The quota is controlled by the institution which can raise or lower the cut-off score required for admission.

The two column sums $b^+$ and $b^-$ are the "base-rates" that characterize the applicant pool. As $b^+$ tends towards 1, it will eventually surpass the total proportion of correct decisions, prc = tn + tp, for any fallible test. At this point, the gain in correct decisions, prc-$b^+$, becomes negative. In other words, the test becomes worse than useless if one is interested in minimizing misclassification rates, because their sum is smaller for purely random admissions. Readers interested in 1) the relationship between hit rates and correlational test validity for strictly binary criteria and in 2) lower limits for correlations necessary to improve over random admissions in overall percent correct decisions might consult Schonemann and Thomson (1996, p. 10) and Schonemann (1997, p. 191), respectively.

As Table 5B shows, this event arises with disconcerting frequency for tests with validities near .15, such as the GRE (next to last column), as long as there are more qualified than unqualified candidates. For base-rates surpassing .50 by as little as .05, the gain in correct classifications over random admissions tends already to be negative for low quotas. Once the positive base-rate reaches .6, random admissions are uniformly superior to a test with validity .15. If the positive base-rate exceeds .7, the gains are negative across all quotas (with one exception, .01), even for validities as large as .5. This is the reason why it matters whether the validities in the recent GRE Report are artificially inflated as a result of the flawed PDA estimation method.

Readers interested in the relation of this approach to Signal Detection Theory and in any further details which otherwise would lead astray from the main theme may be referred to Schonemann and Thompson (1996) and Schonemann (1997, 2005). The main point here was simply to note that the conventional test theory approach is inconclusive and that a more direct assessment of the practical utility of tests, which dispenses with the need for questionable post hoc corrections, is available if desired.

*Test Bias*

Instructive applications of this decision-oriented point of view bearing on the problem of test bias can be found in Crouse and Trusheim (1988, p. 104). They concluded: "The SAT acts much like a supplement to high school rank with zero validity that reject additional blacks" (p. 107). For B.A. as a criterion, they found that adding the SAT to High School Rank increases the proportion of false negatives for Blacks from .17 to .28, while those for the Whites remained the same (.13). On the basis of such results they concluded that adding the SAT to high school record increases the bias against Blacks (Chapter 5: The SAT has an adverse impact on Black applicants). The same is true for low-income applicants (Chapter 6). Schonemann and Thompson (1996) confirmed this fact for other data sets.

The traditional treatment of the bias problem in regression terms (Cleary, 1968) tends to obfuscate this socially relevant fact. Yet ETS releases and APA publications typically promote Cleary's definition of "test bias", which converges on the absurd conclusion that tests, if biased at all, are biased *in favor* of minorities. The same publications usually avoid mention of Cole's (1973) definition, which implies that tests are biased *against* minorities. This is to be expected if only because minorities are less likely to be able to afford the stiff coaching fees for retaking such tests if they do not succeed on first try. This type of ("hit-rate") bias becomes apparent to the naked eye by simply inspecting Crouse and Trusheim's data in chapters 5 and 6. Hartigan and Wigdor (1988, p. 260) concluded in their critique of Hunter's exaggerated claims of the monetary benefits of mental tests (US Department of Labor, 1983):

"At this point in history, it is certain that the use of the GATB without some sort of score adjustment would systematically screen out blacks, some of whom could have performed satisfactorily on the job. Fair test use would seem to require at the very least that the inadequacies of the technology should not fall more heavily on the social groups already burdened by the effects of past and present discrimination" (p. 260).

Though aimed at a different test, the same considerations apply to the SAT and GRE.

*Conclusion*

In summary it can be stated that the GRE validities presented in the GRE Report are result from the bias-enhancing pooling method PDA whereby the positive bias increases as population validities decrease. Furthermore, correlational validity tells us nothing about the merit of a test since its practical validity is a joint function of validity, base rate, and admission quota. Although this has been known since Meehl and Rosen (1955) this point went largely unheeded.

Table 1

*Excerpts from Tables C1, C2, and C6 of the B & W Report*

*A. Correlations (C2, p.56. Pooled R's from C1, p. 55)*

| Dept. | A | B | C | D | E | | | | | |
|---|---|---|---|---|---|---|---|---|---|---|
| | | | | | | *n*'s | | | pooled *R*'s | |
| Biology | | | | | | | | | | |
| SSS (N's) | 15 | 38 | 10 | 58 | 24 | 145 | U | | VQ | UVQ |
| R | .363 | .540 | .607 | .274 | .448 | | | .22 | .33 | .40 |
| Rc | .84 | .813 | .739 | .325 | .537 | | | .34 | .51 | .57 |
| Chemistry | | | | | | | | | | |
| SSS (*n*'s) | 85 | 49 | | | | 134 | | | | |
| R | .436 | .281 | | | | | | .28 | .36 | .46 |
| Rc | .567 | .498 | | | | | | .45 | .5 | .62 |
| Education | | | | | | | | | | |
| SSS (*n*'s) | 138 | 453 | 108 | | | 701 | | | | |
| R | .486 | .325 | .479 | | | | | .29 | .29 | .38 |
| Rc | .497 | .392 | .578 | | | | | .35 | .32 | .44 |
| English | | | | | | | | | | |
| SSS (*n*'s) | 45 | 62 | 19 | 34 | 10 | 175 | | | | |
| R | .678 | .068 | .6 | .385 | .909 | | | .11 | .39 | .4 |
| Rc | .762 | .088 | .663 | .501 | .972 | | | .16 | .45 | .47 |
| Psychology | | | | | | | | | | |
| SSS (*n*'s) | 52 | 41 | 13 | 49 | | 155 | | | | |
| R | .441 | .256 | .469 | .504 | | | | .16 | .37 | .41 |
| Rc | .536 | .424 | .64 | .695 | | | | .29 | .51 | .57 |

*Note*: the entries in the body of the table refer to UVQ as predictors.

*B. Numerical illustration of the PDA method of pooling:*

*Chemistry:[85(.436)+49(.498)]/134 = .459.*

*C. Regression weights for Chemistry*
*(predicting cumulative graduate GPA from UVQ (C2, p.56)*

| | | |
|---|---|---|
| SSS | 85 | 49 |
| U | .298 | .147 |
| GRE-V | .213 | -.069 |
| GRE-Q | -.013 | .389 |

*Table 2*

*General Case*

*1. Population correlation matrix:*

|   | 1 | 2 | 3 |
|---|---|---|---|
|   | 1 | .600 | .200 |
|   | .600 | 1 | .300 |
|   | .200 | .300 | 1 |

*2. Sample multiple correlations followed by bias:*

| NSS | SSS |  | 1 | 2 | 3 | diff |
|---|---|---|---|---|---|---|
| **40** | **25** | **pop** | **.600** | **.627** | **.301** |  |
|  |  | pda | .612 | .645 | **.369** |  |
|  |  | agr | .601 | .636 | **.336** |  |
|  |  | sum |  |  | **.311** |  |
| **Bias:** |  |  |  |  |  |  |
|  |  | pop-pda | -.011 | -.017 | **-.068\*\*** |  |
|  |  | pop-agr | -.001 | -.009 | **-.035** | **.033** |
|  |  | pop-sum |  |  | **-.010\*** |  |
| **20** | **50** | **pop** | **.600** | **.627** | **.301** |  |
|  |  | pda | .652 | .657 | **.345** |  |
|  |  | agr | .646 | .657 | **.325** |  |
|  |  | sum |  |  | **.321** |  |
| **Bias:** |  |  |  |  |  |  |
|  |  | pop-pda | -.052 | -.030 | **-.044\*\*** |  |
|  |  | pop-agr | -.046 | -.029 | **-.024** | **.020** |
|  |  | pop-sum |  |  | **-.020\*** |  |
| **13** | 77 | pop | .6 | .627 | **.301** | |
| | | pda | .639 | .669 | **.363** | |
| | | agr | .635 | .666 | **.353** | |
| | | sum | | | **.331** | |
| **Bias:** | | | | | | |
| | | pop-pda | -.039 | -.042 | **-.062**** | |
| | | pop-agr | -.035 | -.039 | **-.052** | .01 |
| | | pop-sum | | | **-.030*** | |

*Table 3*

*Zero Validities*

*1. Population correlation matrix:*

|   | 1 | 2 | 3 |
|---|---|---|---|
| 1 | 1 | .600 | .000 |
| 2 | .600 | 1 | .000 |
| 3 | .000 | .000 | 1 |

*2. Sample multiple correlations followed by bias:*

|  | NSS | SSS |  | 1 | 2 | 3 | diff |
|---|---|---|---|---|---|---|---|
| **40** |  | **25** | **pop** | **.600** | **.600** | **.000** |  |
|  |  |  | pda | .612 | .615 | **.270** |  |
|  |  |  | agr | .591 | .590 | **.040** |  |
|  |  |  | sum |  |  | **.022** |  |
| **Bias:** |  |  |  |  |  |  |  |
|  |  |  | pop-pda | -.012 | -.015 | **-.270\*\*** |  |
|  |  |  | pop-agr | .009 | .010 | **-.040** | **.230** |
|  |  |  | pop-sum |  |  | **-.022\*** |  |
|  |  |  |  |  |  |  |  |
| **20** |  | **50** | pop | .600 | .600 | **.000** |  |
|  |  |  | pda | .605 | .602 | **.163** |  |
|  |  |  | agr | .597 | .598 | **.025** |  |
|  |  |  | sum |  |  | **-.010** |  |
| **Bias:** |  |  |  |  |  |  |  |
|  |  |  | pop-pda | -.005 | -.002 | **-.163\*\*** |  |
|  |  |  | pop-agr | .003 | .002 | **-.025** | **.138** |
|  |  |  | pop-sum |  |  | **.001\*** |  |
|  |  |  |  |  |  |  |  |
| **13** |  | **77** | pop | .600 | .600 | **.000** |  |

|   |         |       |       |          |       |
|---|---------|-------|-------|----------|-------|
|   | pda     | .636  | .637  | **.156** |       |
|   | agr     | .63   | .631  | **.029** |       |
|   | sum     |       |       | **-.014**|       |
| **Bias:** |   |       |       |          |       |
|   | pop-pda | -.036 | -.037 | **-.156\***| |
|   | pop-agr | -.030 | -.031 | **-.029**| **.127** |
|   | pop-sum |       |       | **.014\***|     |

*Table 4*

*Bias as a function of validity*

1. Population correlation matrices:

| | | | | | | | | | | | | | | | |
|---|---|---|---|---|---|---|---|---|---|---|---|---|---|---|---|
| | 1 | .6 | 0 | 1 | .6 | .1 | 1 | .6 | .1 | 1 | .6 | .2 | 1 | .6 | .4 |
| | .6 | 1 | 0 | .6 | 1 | .1 | .6 | 1 | .2 | .6 | 1 | .3 | .6 | 1 | .2 |
| | 0 | 0 | 1 | .1 | .1 | 1 | .1 | .2 | 1 | .2 | .3 | 1 | .4 | .2 | 1 |
| **validity** | .000 | | | .112 | | | .202 | | | .301 | | | .403 | | |

2. Bias (row/col means: of magnitude):

| Pop. validity | NSS | 40 | 20 | 13 | | |
|---|---|---|---|---|---|---|
| | SSS | 25 | 50 | 77 | rmns | diff |
| | pda | -.270 | -.163 | -.156 | **.196** | |
| **.000** | agr | -.040 | -.025 | -.029 | **.031** | **.165** |
| | sum | -.022 | .001 | .014 | **.012** | |
| | pda | -.160 | -.068 | -.040 | **.089** | |
| **.112** | agr | .001 | -.011 | -.008 | **.007** | **.082** |
| | sum | .007 | -.008 | -.007 | **.007** | |
| | pda | -.094 | -.026 | -.053 | **.058** | |
| **.202** | agr | .016 | .031 | .009 | **.019** | **.039** |
| | sum | .108 | .072 | .031 | **.070** | |
| | pda | -.088 | -.069 | -.031 | **.063** | |
| **.301** | agr | -.036 | -.037 | -.020 | **.031** | **.032** |
| | sum | .031 | -.008 | -.020 | **.020** | |
|       | pda | -.027 | .013 | -.017 | **.019** |      |
| **.403** | agr | .009  | .021 | -.011 | **.014** | **.005** |
|       | sum | .105  | .068 | .024  | **.047** |      |
|       | cmns| .080  | .041 | .031  | .051 |      |

*Table 5*

*A. Assessing Test Utility in Terms of Classification Rates*

When there are exactly two outcomes (e.g., gr, will graduate, and ngr, will not graduate), and two decisions based on the test score (e.g., pass = 1, fail =0 ), then the four possible outcomes can be summarized in a fourfold table of joint proportions of the form

|  |  | **Criterion: Graduation** |  |  | **Numerical Illustration** |  |  |
|---|---|---|---|---|---|---|---|
|  |  | ngr | gr |  | $r = .15$ |  |  |
|  |  | - | + | sum | - | + | sum |
|  |  |  | q |  |  |  |  |
| Predictor: Test Results | + | fp | tp | + | .1 | .2 | .3 |
|  | - | tn | fn | - | .3 | .4 | .7 |
| sums |  | b$^-$ | b$^+$ |  | .4 | .6 | 1 |

The four cells in the body of the tables contain the proportions of the four possible joint outcomes:

tp (= .2, "true positives") proportions of applicants who pass the test and would graduate, if admitted tn (= .3, "true negatives" proportion of applicants failing the test and would not graduate, if admitted fp (= .1, "false positives" proportion of applicant passing the test, but would not graduate, if admitted fn (= .4, proportion of applicants failing the test who would graduate, if admitted.

The marginal row sum, q = fp+tp is the (admission-) "quota (= .3), i.e., the proportion passing the test. In practice the quota is set by the institution that chooses a suitable cut-off for admission. Selective schools (e.g., Yale) choose higher cut-offs, resulting in smaller quotas, than less selective schools.

The two column sums, $b^-$ and $b^+$ are called negative (resp. positive) "base-rates". The positive base-rate (.60) is the proportion of applicants who would graduate, if they were admitted. Typically, $b^+$ exceeds .5, and the current discussion will be limited to this case.

The sum of the diagonal elements, tp+tn (= .5) is the "total proportion of correct decisions", prc. The difference,

gain = prc-$b^+$

expresses the gain (or loss, if it is negative) in correct classification as a result of using the test, compared to the proportion of correct decisions achievable without the test by adopting a random admission policy (which reflects the base-rates into the admission sample).

Finally, the ratio hr = tp/$b^+$ is the "hit-rate" of the test. It estimates the probability that a qualified candidate (i.e., a candidate who would graduate, if admitted) will pass the test. Occasionally these proportions are multiplied by 100 to express them as percentages.

**Numerical Illustration:**

In the above numerical example, the validity (tetrachoric correlation) is .15, corresponding to the validity of the GRE for cumulative graduate GPA. The positive base-rate is .6 and the admission quota is .3. The total proportion of correct classifications on using the test is

prc = tp+tn = .2+.3 = .5. Random admission would result in a proportion of $b^+$ = .6 of correct classification, because a random admission sample contains 60% qualified candidates. Hence the gain in this case is negative, .5-.6 = -.1, which means that use of the test results fewer correct classifications than random admissions.

The hit-rate is hr = tp/$b^+$ = .2/.6 = .33. This means that only one out of three qualified candidates passes this test, because its validity is so low.

If $b^+ < b^-$, the gain would be defined as gain = prc – $b^-$, that is, more generally, gain = prc – max($b^+$, $b^-$). However, if $b^-$ > .5, the decision rationale would be less convincing, because in this case the optimal decision would be to act as if all candidates are unqualified which, in practice, is unrealistic. A cursory survey of the literature suggests that $b^+$ > .5 is the more typical case.

B. Gains (losses, if negative) in Correct Classifications As a Result of Test Use, Relative to Random Admissions for positive base-rates $b^+ = .5$. %C total percent correct, G/L gains/losses, HR hit rates)

|  |  | $r = .50$ | | | $r = .30$ (SAT) | | | $r = .15$ (GRE) | | |
|---|---|---|---|---|---|---|---|---|---|---|
| $b^+$ | q | %C | G/L | HR | %C | G/L | HR | %C | G/L | HR |
| 50 | 70 | 64 | 40 | 84 | 58 | 8 | 78 | 54 | 4 | 60 |
|  | 60 | 66 | 16 | 76 | 60 | 9 | 70 | 55 | 4 | 65 |
|  | 50 | 67 | 17 | 67 | 60 | 10 | 60 | 55 | 5 | 55 |
|  | 40 | 66 | 16 | 56 | 60 | 9 | 50 | 55 | 5 | 45 |
|  | 30 | 64 | 14 | 44 | 59 | 9 | 38 | 54 | 4 | 34 |
| 55 | 70 | 67 | 12 | 83 | 60 | 5 | 77 | 56 | 1 | 74 |
|  | 60 | 67 | 12 | 74 | 60 | 5 | 68 | 56 | 0 | 64 |
|  | 50 | 67 | 11 | 65 | 60 | 4 | 58 | 54 | -1 | 54 |
|  | 40 | 65 | 10 | 55 | 54 | 4 | 48 | 54 | -1 | 44 |
|  | 30 | 62 | 7 | 43 | 56 | 1 | 38 | 52 | -3 | 34 |
| 60 | 70 | 68 | 9 | 82 | 62 | 2 | 77 | 58 | -2 | 73 |
|  | 60 | 68 | 8 | 73 | 61 | 1 | 67 | 56 | -3 | 64 |
|  | 50 | 66 | 6 | 64 | 60 | 0 | 58 | 54 | -5 | 54 |
|  | 40 | 63 | 3 | 53 | 57 | -3 | 48 | 52 | -8 | 44 |
|  | 30 | 60 | 0 | 41 | 54 | -6 | 36 | 50 | -10 | 34 |
| 65 | 70 | 70 | 5 | 81 | 64 | -1 | 76 | 60 | -5 | 73 |
|  | 60 | 68 | 3 | 72 | 62 | -3 | 66 | 57 | -8 | 63 |
|  | 50 | 65 | 0 | 62 | 59 | -6 | 57 | 55 | -10 | 54 |
|  | 40 | 62 | -3 | 52 | 55 | -10 | 46 | 51 | -14 | 43 |
|  | 30 | 57 | -8 | 39 | 51 | -14 | 36 | 48 | -18 | 33 |
| 70 | 70 | 72 | 1 | 80 | 66 | -4 | 66 | 62 | -8 | 73 |
|  | 60 | 68 | -2 | 70 | 63 | -8 | 67 | 58 | -12 | 63 |
|  | 50 | 64 | -6 | 60 | 58 | -12 | 56 | 54 | -16 | 53 |
|  | 40 | 59 | -11 | 49 | 54 | -16 | 46 | 50 | -20 | 43 |
|  | 30 | 53 | -17 | 38 | 49 | -21 | 35 | 45 | -24 | 33 |

*Note*: Decimal points omitted. Multiply with .01

*Acronyms and Glossary*

A - analytical part of the GRE
AGR - aggregated (pooled) subsamples
$B^+$ - (positive) base rate: proportion of qualified candidates in the unselected population
Correction for attenuation - (up) correction for unreliability
Correction for restriction of range - (upward) correction for selection
Correction for shrinkage - (downward) correction for capitalization on chance
Diff - difference
Distal - long-term
FP - false positives
FN - false negatives
GPA - grade point average
GRE - Graduate Record Exam
Hit Rate - proportion of qualified candidates who pass the test
NSS - number of subsamples
PDA - pooled department analysis (numerical illustration Table 1B)
POP - population
PRC - total percent correct
Proximal - short-term
Q - quantitative part of the GRE, also (admission) quota
R - multiple correlation
Rc - corrected multiple correlation
SAT - Scholastic Aptitude (Achievement) Test
SSS - subsample size
SST - total sample size
SUM - unweighted sum of predictors
teens - validity between .10 and .19
TP - true positives
TN - true negatives
U - undergraduate record
V - verbal part of the GRE